\documentclass[aps, preprint, showpacs, epsfig]{revtex4}
\usepackage{graphics}
\usepackage{graphicx}
\usepackage{epsfig}
\def \be {\begin{equation}}
\def \ee {\end{equation}}
\def \ba {\begin{eqnarray}}
\def \ea {\end{eqnarray}}
\def \bm {\begin{displaymath}}
\def \em {\end{displaymath}}

\begin{document}
\title{ Shear unzipping of double stranded DNA}
\author{Shikha Prakash and Yashwant Singh }
\affiliation{Department of Physics, Banaras Hindu University,
Varanasi-221 005,
India}

\date{\today}
\begin{abstract}
We use a simple nonlinear scaler displacement model to calculate
the distribution of effect created by a shear stress on a double stranded
DNA (dsDNA) molecule and the value of shear force $F_c$ which is required to separate 
the two strands of a molecule at a given temperature. It is shown that for molecules of
base pairs less than 21, the entire single strand moves in the direction of applied
force whereas for molecules having base pairs more than 21, part of the strand
moves in the opposite direction under the influence of force acting on the other strand.
This result as well as the calculated values of $F_c$ as a function of length
of  dsDNA molecules are in very good agreement with the experimental
values of Hatch et al. (Phys. Rev. E $\bf 78$, 011920 (2008)).
 
\end{abstract}

\maketitle
\section*{ I. INTRODUCTION}
A double stranded DNA (dsDNA) molecule consists of two polynucleotide
strands connected loosely by hydrogen bonds through the base pairs
and base-stacking between nearest neighbour pairs of base pairs 
and wound around each other to make a helix. The constraints of this 
helical structure require that the two base sequences on opposite
strands must be complementary, with Adenine (A) always binding to
Thymine (T) and Guanine (G) binding to Cytosine (C) \cite{saenger}. 
The force that holds the complementary strands of DNA together 
is an important regulator of life's processes because the
binding of regulatory proteins to DNA often involves the 
procedure of mechanical separation of its strands. 
The intermolecular forces of DNA have been studied extensively
using a variety of techniques,  which cover a broad range of forces
from a few piconewton(pN) up to several hundred piconewtons (pNs). 

These techniques use either atomic force microscopy (AFM)
or laser optical
traps and  magnetic tweezers \cite{strick,bockel,smith,reif,conroy,kumar}. 
The experiments
fall into two general categories; those conducted on short
DNA and those on long DNA. Experiments on long DNA focussed on the overall
properties of the molecule \cite{bockel,zhang} and resulted in the discovery of 
S-DNA \cite{reif,bloom,clausen,busta}; the occurance of this so called B-S transition
has also been found in a short DNA chain of 30 base pairs \cite{morfill}.
 Those on short duplexes focussed on the reaction
pathway of melting, and contributed to the understanding of the local
unbinding of DNA \cite{lee,kerall,pope,strunz,kuhner}. In the single molecule experiments,
the result may depend on choice of which variables are fixed and which
can fluctuate. In some of the experiments (like AFM) extension 
is fixed and force is allowed to fluctuate (constant extension ensemble)
while in others (like magnetic tweezers) the force is fixed and extension
is allowed to fluctuate (constant force ensemble). 

When a force is applied to pull apart the two strands from one end
of a dsDNA molecule in a direction perpendicular to the helical 
axis, bases are sequentially stretched as the duplex is unzipped.
On the other hand, in shear unzipping in which the applied force pulls the two strands
from opposite ends as shown in Fig.1, the stretching is spread
out over many base pairs. The study of distribution of shearing 
force along the length of a DNA molecule may lead to valuable 
informations about the force distribution across the phosphate
backbone of a single strand of DNA versus the force distribution 
across the paired bases of complementory DNA. Such informations
are of relevance not only to understanding the biological processes
but also in material science application such as determining 
the strength of DNA/gold nanoparticle assemblies \cite{chak}.

Following the work of Lee et al. \cite{lee} many workers 
\cite{pope,strunz,sattin,grange,schu,neurt,milam,wal,noy,hatch} measured
the value of shear force which separates the two strands of a dsDNA molecule.
It was found that the  force increases linearly
with the length. Using a simple ladder model of DNA and expressing
the backbone bond energy as well as the interaction energy among complementary
bases in the form of harmonic springs, de Gennes \cite{gennes} in 2001
predicted that the critical force for shear unzipping which for short 
chains shows linear dependence would saturate to a finite value in the
limit where the number of base pairs approaches infinity and that the shear
stress relaxes over a distance  (number of base pairs) $\chi^{-1}$  (=$\sqrt\frac{\kappa}{2R}$,
where $\kappa$  is the spring constant characteristic of stretching the backbone,
and R is the spring constant characterstic of stretching the hydrogen bonds 
between base pairs) on either side of the chain. The critical force $F_c$
for shear unzipping as a function of number of bound base pairs was given
\cite{gennes} as 
\be
F_{c}=2f_{c}[\chi^{-1}tanh(\chi\frac{N}{2})+1].
\ee
where $f_c$ is the rupture force of a single bond and (N+1) is the 
number of base pairs in the molecule.

Chakrabarti  and Nelson \cite{chak} have generalized the simple
harmonic model of deGennes \cite{gennes} by representing the interaction
among complimentary bases by the Lennard-Jones (12-6) potential model
and found that the strain is indeed localized over a narrow range of $\chi^{-1}$
on either side of the chain and the chain unzips when the force exceeds 
a critical value of $F_c\sim f_cN$ for short chains and $F_c\sim 2f_c\chi^{-1}$
for long ones as predicted by deGennes \cite{gennes}. It may, however,
be noted that the calculations of 
deGennes \cite{gennes} and also of Chakrabarti and Nelson \cite{chak} correspond to zero
temperature. The Langevin dynamics simulation has recently been used to study
shear unzipping of dsDNA at finite temperature \cite{mishra}. The results found
for $F_c$ are in agreement with the results of deGennes \cite{gennes} and Chakrabarti
and Nelson \cite{chak}.

Hatch et al. \cite{hatch} have measured the value of $F_c$ for several dsDNA 
molecules of length ranging from 12 to 50 base pairs at the room temperature 
and found that $F_c$ is linear function of molecular length only up to 20 base pairs
and approaches to an asymptotic value as the number of base pairs increases. 
In fact they found that $F_c$ for a molecule of 32 base pairs has already reached 
within $5\%$ of the asymptotic value which was found to be 61.4 pN.
But when Hatch et al. \cite{hatch} tried to fit their data to Eq.(1)
they found that this can be done only by assuming 7 base pairs of all
chains to be in the open state, i.e. the effective length of a molecule 
of (N+1) base pairs is (N-6) irrespective of the value of N. They attributed 
this to temperature effect as deGennes calculation did not include temperature. 

In this paper we  calculate the value of $F_c$ as a function of the number
of base pairs at room temperature and compare our results with the experimental
data of Hatch et al. \cite{hatch} and show that the need to adjust the length 
of a molecule is not as much  due to temperature as due to use of the  values of $f_c$
and $\chi$ in Eq.(1) and the nature of the curve of $F_c$ $\it vs$ N. We derive another 
form of Eq.(1) and show that with 
reasonable choice of values of $f_c$ and $\chi$ one gets values of
$F_c$ which are in better agreement with experiment without adjusting 
the length of molecules than that found from Eq.(1). The model which we describe 
in Sec. II is similar to the nonlinear 
scaler displacement model of Chakarbarti and Nelson \cite{chak}.
We calculate the force in both the constant force and constant
extension ensembles. The paper is organized as follows.
In Sec II we describe the model and calculational procedures. In Sec III
we give results found for distribution of effects created by  
shear force along the length of a molecule and the value of 
$F_c$  as a function of number of base pairs. In 
Sec IV we compare our results with those of deGennes \cite{gennes} and with
the experimental values \cite{hatch}. The paper concludes with a brief
comment given at the end of Sec IV.

\section*{II. MODEL}
We consider a dsDNA molecule of length (N+1) base pairs which 
both 5$^\prime$-ends (or both 3$^\prime$-ends) are pulled along the helical
(molecular) axis
by a force $\bf F$ as shown in Fig.1. The displacements of $\it i^{th}$
nucleotide from its equilibrium position are denoted by $\bf {u}_i$
for one (the lower one in Fig.1) single stranded DNA (ssDNA) chain and by $\bf {v}_i$ for the
other (the upper one) chain. The effective Hamiltonian of the
system can be written as \cite{chak}
 
\ba
H=&&\sum_{i=-N/2}^{N/2-1}\frac{1}{2}\kappa\large[({\bf u}_{i+1}  
-{\bf u}_{i})^2 +({\bf v}_{i+1} -{\bf v}_{i})^2\large]
+\sum_{i=-N/2}^{N/2}v(|{\bf u}_{i} -{\bf v}_{i}|)\nonumber \\
 &&-{\bf {F}}.({\bf{u}}_{N/2}-{\bf{v}}_{-N/2}).
\ea  

The first term of this equation represents the stretching 
energy of nucleotides, excecuting a simple harmonic motion
with a spring constant $\kappa$  along each chains in dsDNA. 
In the second term, $v(|\bf{u}_i-\bf{v}_i|)$ represents the potential
energy of interaction between bases in the $\it {i}^{th}$ pair 
and the last term of Eq.(2) represents the energy contribution due to
the shear stress. In writing Eq.(2) we did not include  the contribution
arising due to the helicity of DNA as this effect was  found to be negligible
by Hatch et al. \cite{hatch} and also by Lavery and Lebrun \cite{lavery}.

For the potential $v(|\bf{u}_i-\bf{v}_i|)$ we use a simple model
which has a hard-core repulsion and a long range attraction,

$\hspace{5cm}v(|{\bf u}_{i} -{\bf v}_{i}|)=\infty
\hspace{4cm}for \hspace{2mm} 
\frac{|{\bf z}_i|}{\sigma}< 0, $

\be
\hspace{6.5cm}=-\frac{\epsilon}{\left( 1+\frac{|\bf{z}_i|}{\sigma}\right)^6} 
\hspace{2cm}for\hspace{2mm}\frac{|{\bf z}_i|}{\sigma}> 0.
\ee

Here $|\bf{z}_i|$ is the magnitude of increase in length of hydrogen
bonds connecting bases in the $\it{i}^{th}$ pair  from its equilibrium 
value, $\epsilon$ is the depth of potential at the equilibrium separation
and $\sigma$ is the diameter of dsDNA (see Fig.2).  The repulsion represents 
the steric hindrance 
which forbids the molecule from getting compressed along the bond 
linking the bases with respect to its equilibrium value. 
From Fig.2 one finds ,

\be
\left( 1+\frac{|\textbf{z}_i|}{\sigma}\right)= \left(1+\frac{(|{\bf u}_{i} -{\bf v}_{i}|)^2}{\sigma^2}\right)^{1/2},
\ee

We take the value of $\sigma$ equal to $20\rm \AA$ which is the diameter 
of the {\it Canonical} DNA (B-DNA) at  room temperature.

Since in shear unzipping, a molecule is stretched along its axis we consider
the longitudinal displacements of nucleotides and neglect 
the transverse displacements; the transverse displacements have been 
found in ref \cite{chak} an order of magnitude smaller than those in the direction 
of shear. We define new variables,

\be
x_{i}=\frac{u_{i}+v_{i}}{\sqrt{2}},\hspace{1cm} y_{i}=\frac{u_{i}-v_{i}}{\sqrt{2}},
\ee

where, $u_i$ and $v_i$ now represent the longitudinal
displacements of respective strands. In the notations used here $u_i$ and $v_i$
are positive when the $i^{th}$ nucleotide of upper and lower strands move to {\it r.h.s.}
and negative when they move to {\it l.h.s.}. For the experimental situation
shown in Fig. 1 it is clear that for $i>0$, $u_i>0$ and magnitute of $u_i$ is greater
than that of $v_i$ and for $i<0$, $v_i<0$ and the magnitute of $v_i$ is greater 
than that of $u_i$. This leads to following relation for variables $x_i$ and
$y_i$;

\be
x_{-i}=-x_{i},\hspace{2mm} x_{0}=0\hspace{2mm} and \hspace{2mm}y_{-i}=y_{i}.
\ee

When we substitute these variables in Eq.(2) it decouples into two 
independent components;

\be
H=H_{x}+H_{y},
\ee

where

\be
H_{x}=\sum_{i=-N/2}^{N/2-1}\frac{1}{2}\kappa(x_{i+1}-x_{i})^2
-\frac{F}{\sqrt{2}}(x_{N/2}-x_{-N/2}),
\ee
and

\be
H_{y}=\sum_{i=-N/2}^{N/2-1}\frac{1}{2}\kappa(y_{i+1}-y_{i})^2
+\sum_{i=-N/2}^{N/2}v(y_{i})-\frac{F}{\sqrt2}(y_{N/2}+y_{-N/2}).
\ee

Here the potential $v(|\bf{u}_i-\bf{v}_i|)$ defined in Eq.(3) is expressed
in terms of variable $y_i$. Using Eqs. (4) and (5), one can rewrite Eq.(3) as

\be
v(y_{i})=-\frac{\epsilon}{\left(1+\frac{2y_{i}^2}{\sigma^2}\right)^3}.
\ee

Note that the expression of $H_x$ does not contain the on-site
potential $v(y_i)$ and simply corresponds to a harmonic chain
which is being pulled at the two ends by a force $F/\sqrt 2$
whereas the expression of $H_y$ contains the on-site potential
$v(y_i)$ as well as the force term.

In view of the relations given by Eq.(6) the average value of 
displacement  $<x_n>$ of  $\it {n}^{th}$ base pair can be calculated
from the relation,

\ba
<x_{n}>=\frac{ \int \prod_{i=0}^{N/2}dx_{i} x_{n}\exp(-\beta H_{x})}
{ \int \prod_{i=0}^{N/2}dx_{i} \exp(-\beta H_{x})},
\ea
where $\beta=(k_BT)^{-1}$, $k_B$ being the Boltzmann constant and 
$T$ is the temperature. Since for $H_x$ given by Eq.(7) the integrals
in Eq.(11) are Gaussians, one can solve them analytically to give
\be
<x_{n}>=\frac{F}{\sqrt{2}\kappa}n.
\ee

This result agrees with the one found by deGennes \cite{gennes}.

The average value of displacement $y_n$ of $\it{n}^{th}$
base pair can be found from the relation,
\be
<y_{n}>=\frac{ \int \prod_{i=-N/2}^{N/2}dy_{i} y_{n}\exp(-\beta H_{y})}
{ \int \prod_{i=-N/2}^{N/2}dy_{i} \exp(-\beta H_{y})}.
\ee

The integrals appearing in this expression cannot be evaluated 
analytically because of the form of on-site potential $v(y_i)$.
However, for the expression of $H_y$ given by Eq.(9) the integral 
appearing in Eq.(13) reduces to multiplication of ($N+1$) matrices.
The discretization of the coordinate variable and introduction of a proper
cut-off on the maximum values of $y's$ determine the size of the
matrices. We have taken -40$\rm \AA$ and 40$\rm \AA$ as the
lower and upper limit of integration for each co-ordinate
variable and discretized space using the Gaussian-Legendre 
method with the number of grid points equal to 900. By changing the
limits of integration as well as the number of grid points we made it 
sure that the values of $<y_n>$ are independent of the limit
of integration and the number of grid points chosen to discretize 
coordinate variable.

\section*{III. RESULTS}
When a shear force is applied on a dsDNA molecule, its two ssDNA strands 
get pulled in opposite directions as shown in Fig.1. The bonds in the backbone
of DNA as well as bonds connecting bases in a pair are stretched.
In a case of $\kappa$ being infinitely large the two strands will
move like a rigid body pulling all base pairs in the sequence in parallel.
The effect of the shear force will then be uniformly distributed
across all of the base pairs. However, if $\kappa$ is finite, then
both the backbone and the base pairs will stretch when a shear force
is applied. The effect of the shear force may then be confined 
to limited lengths on both ends of the molecule. To see
how the effect caused by shearing of a dsDNA molecule
is distributed along the length of a molecule and how this depends on 
the energies associated with the stretching
of backbone and base pairs we calculate the value of 
$<y_n>$ from Eq.(13) for different values of $n$ when the two 
end base pairs are stretched to a given length by the shear force.

In Fig.3 we plot our results for four dsDNA molecules of length
17, 25, 33 and 49 base pairs and for $y_{-N/2}=y_{N/2}=1.50\rm\AA$,
$2.0\rm\AA$, $2.38\rm\AA$ and $2.60\rm\AA$. The values shown in the figure correspond
to $\epsilon=0.04eV, \kappa=0.10eV/\rm\AA^2$ and $T=300^0K$. When
$y_{-N/2}$ and $y_{N/2}$ are allowed to be free (i.e.$\bf{F}=0$),
then $<y_n>$ is found to be zero for all $n$. From Fig.3 it is clear
that for short chains the effect created by shear force is 
distributed along the entire length of a molecule affecting all base pairs,
whereas, for relatively larger molecules the base pairs in the central
part of molecules are only marginally affected. For example, for a molecule
 of length 17 base pairs the value of $<y_0>$ is 1.76$\rm\AA$
when $y_{-8}=y_8=2.38\rm\AA$, whereas, for the similar situation 
(i.e. $y_{-24}=y_{24}=2.38\rm\AA$) the value of $<y_0>$ for a molecule
of length 49 base pairs is only 0.27$\rm\AA$. The qualitative nature of these
 results are in agreement 
with the results found in refs \cite{chak} and \cite{mishra}. The other point to be noted
from the figure is that the qualitative nature of the distribution 
of the effect of shear stretching along length of a molecule 
is same for all values of $y_{-N/2}=y_{N/2}$ plotted in the figure.
  
In Fig.4 we plot the value of $<y_0>$ when $y_{-N/2}=y_{N/2}$ = 2.38$\rm\AA$
as a function of length of dsDNA molecules. As discussed below,
the value of  $y_{-N/2}=y_{N/2}$ = 2.38$\rm\AA$ is assumed to be the critical 
value of stretching in the sense that initiation of separation
of two strands starts at this value of $y_{N/2}$  and the shear force which 
creates this value of stretching of the end base pairs is equal to $F_c$, the
minimum ( or critical) force required to separate the two strands 
of a dsDNA molecule. From Figs. 3 and 4 it is clear that as one moves from
either ends the differential force across the base pairs decreases 
and becomes very small at the centre for larger molecules 
but has not become zero at the centre even for a molecule of 
length 49 base pairs. The extension of the curve of Fig.4 
shows that it would be zero at the centre only for molecules 
of length larger than 80 base pairs.

The average value of  displacements $<u_n>$ and $<v_n>$ of $n^{th}$ base pair
can be found from the known values of $<x_n>$ and $<y_n>$. From Eq.(5) 
one gets
\be
<u_{n}>=\frac{<x_{n}>+<y_{n}>}{\sqrt{2}} \hspace{2mm}and\hspace{2mm}
<v_{n}>=\frac{<x_{n}>-<y_{n}>}{\sqrt{2}}.
\ee
In Fig.5 we plot the values of $<u_n>$ and $<v_n>$ as a function of $n$
for molecules of length 17, 21 and 49 base pairs. The values plotted
in this figure correspond to  $y_{-N/2}=y_N/2$ = 2.38$\rm\AA$ and 
F=40.9, 47.4 and 60.6 pN , respectively for molecules of 
length 17, 21 and 49 base pairs.
These values of force, as is shown below, are the critical force of 
shearing of the respective molecules. From the figure we note that 
while for a molecule of 17 base pairs the entire ssDNA strand
moves in the direction of applied force whereas in the case of a molecule
of 49 base pairs 
nearly half of the strand moves in the opposite direction. While for a molecule of 
 17 base pairs, $<u_{-8}>= 0.66\rm\AA$, $<u_8>= 2.70\rm\AA$  
and $<v_{-8}>= 2.70\rm\AA$, $<v_8> = 0.66\rm\AA$, for a molecule
of 49 base pairs $<u_{-24}>= -2.86\rm\AA$, $<u_{24}>= 6.22\rm\AA$
and $<v_{-24}>= 6.22\rm\AA$, $<u_{24}>= -2.89\rm\AA$. 

As long as the entire ssDNA strand moves in the direction of applied
force the shear force, as explained in ref \cite{chak}, depends linearly 
on molecular length. The departure from the linear dependence of the shear
force on molecular length is expected to take place when $<u_{-N/2}>$ and
$<v_{N/2}>$ become zero i.e. the displacement of nucleotides on the opposite
side of a ssDNA strand remains unaffected by the applied force. Indeed, we find
that $<u_{-10}>$ and $<v_{10}>$=0 for a molecule of length 21 base pairs whcih is in 
very good agreement with the experimental value \cite{hatch}. Since for molecules 
of base pairs larger than 21, part of a ssDNA moves in the opposite direction, a region 
develops inbetween on each strand which remains unaffected by the force. This region moves towards
the centre on increasing the molecular length; the force gets saturated as soon as
the region reaches the centre of the chain and stays there on further increasing 
the molecular length.

The value of shear force needed to separate the two strands of a dsDNA
molecule can be calculated in two different ways. In one, we follow a
method proposed by deGennes \cite{gennes} and which from hereon is referred to 
as a method of constant force ensemble. The other method is based on the constant 
extension ensemble.

In the method of constant force ensemble one first defines a critical
distance $\bar y$ for the rupture of a base pair (i.e. when $y_n$ of 
$n^{th}$  base pair becomes larger than $\bar y$ the bases of the
pair become free) and calculate the value of force which can stretch
a base pair to the critical distance $\bar y$. This force can be 
found from the on-site potential $v(y)$. Thus
\be
f_{c}= -\frac{\partial v(y)}{\partial y}|_{y={\bar y}}\hspace{2mm}
=\frac{12\epsilon \bar y}{\sigma^2} \left(1+\frac{2\bar y^2}{\sigma^2}\right)^{-4}.
\ee
If we take the value of $\bar y=2.38\rm\AA$ ,we find $f_c=4.1pN$
for the value of $\epsilon$ and $\kappa$  given above. This value 
of $f_c$ is close to the value used by Hatch et al \cite{hatch}  to 
fit their experimental data and the value estimated by Chakrabarti and 
Nelson \cite{chak}. To rupture the end base pairs of a given
molecule they must be stretched by the shear force to distance $\bar y$.
The balance of force at one ends of one of the ssDNA gives \cite{gennes};

\be
F_{c}=\kappa(<u_{N/2}>-<u_{N/2-1}>)+f_{c}.
\ee
Using the relations of Eq.(5) and Eq.(12) we get
\be
F_{c}=\sqrt{2}\kappa(<y_{N/2}>-<y_{N/2-1}>)+2f_{c}.
\ee

Taking the value of $<y_{-N/2>}=<y_{N/2}>= \bar y$ =2.38 $\rm\AA$ we calculate
the values of $<y_{N/2-1}>$ from Eq.(13) for several molecules of 
length 10-60 base pairs. The value of force $F_c$ found from Eq.(17)
is shown by dotted line in Fig.6.

In the constant extension ensemble one first calculate the work done in 
stretching the two end base pairs of a given molecule to distance $y$.
This work is found from the relation
\be
W(y)=\frac{1}{\beta}[ln Z_{N+1}(y)-lnZ_{N+1}],
\ee
where
\be
Z_{N+1}(y)=\int\prod_{i=-N/2}^{N/2}dy_{i}\delta
(y_{-N/2}-y)\delta(y_{N/2}-y)\exp[-\beta H_{y}],
\ee
is the constrained partition function and
\be
Z_{N+1}=\int\prod_{i=-N/2}^{N/2}dy_{i}\exp[-\beta H_{y}].
\ee
is the partition function. $\delta$ is the Dirac function.

We used the matrix multipliction method to evaluate $W(y)$
from the above equations for $\epsilon=0.04eV$, $\kappa=0.1eV/\rm\AA^2$
and $T=300^0K$. The derivative of $W(y)$ with  respect  to $y$ gives
the average force $F(y)$ that is needed to keep the extension of 
end base pairs of a given molecule equal to $y$. Thus
\be
F(y)=\frac{\partial W(y)}{\partial y}.
\ee
To get the value of critical force we have to chose a value of $y$
which corresponds to rupturing of base pairs. The values shown
in Fig.6 by dashed line are found when $y$ was taken  equal to $2.0\rm\AA$;
this value is slightly lower than the one taken for the constant force ensemble.
The difference in the value of critical stretching in the two ensembles 
may be due to difference in the path of unzipping.

We note that the values of $F_c$ found by methods of the constant 
force and the constant extension ensembles are close but not identical.
The difference between the two as expected  \cite{swigon} is large for molecules
of smaller lengths but they becomes close as molecular length
increases. Both methods give the same asymptotic value, equal to 61.2pN, which 
is in very good agreement with the experimental value 61.4pN \cite{hatch}.
The experimental values shown in the figure are of Hatch et al \cite{hatch}. 
In view of large spread in experimental data we find good agreement 
between experimental values of $F_c$ and the theoretical values found 
using the constant force ensemble as well as the constant extension ensemble.

\section*{IV DISCUSSIONS }
\subsection*{(a) Comparison with the experiment}
Since the experimental values \cite{hatch} of $F_c$ given in Fig.6 
are found using the fixed force ensemble, we concentrate our discussion
with the theoretical values found using the same ensemble and shown in Fig.6 
by dotted line. The agreement between theory and experiment is excellent 
except for a molecule of length 12 base pairs. We know that at a given 
temperature shorter dsDNA molecules are less stable compared to 
longer molecules. It is quite possible that at room temperature 
due to surface effects few of base pairs at the two ends of a molecule 
of length 12 base pairs are nearly in open state. The curve shown in Fig.6 
by solid line is found when the effective length of a molecule of length
12 base pairs is taken to be equal to 10 base pairs and that of a molecule 
of length 16 base pairs equal to 15 base pairs. The values of force $F_c$ shown by
solid line is in excellent agreement with the experimental values for the entire range
of molecular length investigated by Hatch et al. \cite{hatch}. The saturation 
value of $F_c$ found to be 61.2 pN is also in very good agreement with the experimental 
value of 61.4 pN \cite{hatch}.
\subsection*{(b) Comparison with the results of deGennes \cite{gennes}}
In order to compare our results with those of deGennes \cite{gennes}
we first estimate the value of $\chi$ defined as $\chi=\sqrt\frac{2R}{\kappa}$,
where R is the spring constant of a simple harmonic potential 
between the bases of a pair. Expanding $v(y)$ in ascending powers of $y$ one gets
\be
v(y_{i})= -\epsilon +\frac{1}{2}\left( \frac{12\epsilon}{\sigma^2}\right) y^{2}_i,
\ee
\be
R=\frac{12\epsilon}{\sigma^2}\hspace{2mm}and\hspace{2mm}\chi^{2}
=\frac{24\epsilon}{\kappa \sigma^2}.
\ee
Substituting the values of $\epsilon$, $\kappa$ and $\sigma$ given above 
we find $\chi^{-1}=6.44$ which corresponds to $\kappa/R=83.3$. This value of 
$\kappa/R$ is in good agreement with the predicted value of 77 based on calculation
of the spring constants for base pairs and backbones \cite{value}.

The expression for the displacement $<y_n>$ found by deGennes \cite{gennes} can be written as
\be
<y_{n}>=\frac{1}{2}<y_{0}>(e^{\chi n}+e^{-\chi n}).
\ee
Taking the value of $\chi^{-1}$=6.44 and the values of $<y_0>$ determined above (shown
in Fig 4) we calculate the value of $<y_n>$ from Eq.(24) for molecules of length 
23 and 49 base pairs and compare them in Fig 7 with the values of $<y_n>$ found from
Eq.(13). The two values do not agree; the difference increases with the length of
 molecules. A good agreement is, however, found when $\chi^{-1}$=9.4 is taken. 
This value of $\chi^{-1}$ is about $\sqrt 2$ larger than the value found above.
We calculate $F_c$ from the deGennes equation given by Eq.(1) taking $\chi^{-1}$=6.44
and 9.4 and $f_c$=4.1 pN. The results are plotted in Fig 8 in which we also
plot experimental values and the values found from our approach and shown 
in Fig 6 by dotted line. The values found from Eq.(1) are very different from both 
experimental values and values shown by dotted line. In ref \cite{mishra} the value of 
$F_c/f_c$ as a function of N were found to be in good agreement with the values found
from Eq.(1) for $\chi^{-1}=10$.

Since Eq.(1) involves the rupture force $f_c$ for a base pair and $\chi$,
the value of $F_c$ depends on the values of these quantities. Most often 
quoted value of $f_c$ is in the range of 4-5 pN \cite{chak,hatch} 
and that of  $\chi\sim 0.1$ \cite{chak,hatch,gennes,mishra}. 
Hatch et al. \cite{hatch} took $f_c$=3.9 pN and $\chi^{-1}$=6.8
to fit their data to Eq.(1) and found that they have to adjust the length of molecules
which amounted to shifting the curve of $F_c$ to the right direction by 7 unit of base pairs
in order to get good agreement with the experimental values. We now examine whether such
an adjustment of length is essential or there is some other reason for not getting
good agreement.

In the appendix we derive Eq.(24) from Eq.(9) and show that $\chi$ which appears in Eq.(24)
should be defined as $\chi$=$\sqrt\frac{R}{\kappa}$. Substituting the values of $\kappa$
and R we find $\chi^{-1}$=9.2 which is close to 9.4 found by using the values $<y_n>$
found from Eq.(13). With this definition of $\chi$ we find the following expression
for $F_c$ (see Eq.(A8)).
\be
F_{c}={\sqrt 2}f_{c}[\chi^{-1}tanh(\chi\frac{N}{2})+1].
\ee
The appearance of $\sqrt 2f_c$ instead of $2f_c$ (in Eq.(1)) is due to difference
in the definition of $\chi$. The definition $\chi=\sqrt \frac{R}{\kappa}$ seems
more appropriate in the sense that it is ratio of the two harmonic spring constants 
than the definition $\chi=\sqrt\frac{2R}{\kappa}$ of ref \cite{gennes} in which one spring 
constant is multiplied by two.

The value of $F_c$ found from Eq.(25) when $\chi^{-1}$=9.4 and $f_c$=4.1 pN are compared
in Fig 9 with the values using the method of Sec. III and shown in Fig 6 by dotted
line and values which are found from Eq.(1) when $\chi^{-1}$=6.8 and $f_c$=3.9 pN
\cite{hatch}. From the figure we first note that Eq.(25) gives values of $F_c$ as a
function of molecular length which are close to the values found in
Sec. III (dotted line); the difference between the two is due to the combined
effect of nonlinearity and the temperature. Thus the effect of temperature is not
as large as has been suggested in ref \cite{hatch}. The necessity for 
adjusting the length of molecules
is due to the values of $\chi$ and $f_c$ used in Eq.(1) by Hatch et al. \cite{hatch}.
If one takes $\chi^{-1}$=9.4 and $f_c$=4.1/$\sqrt 2$ pN and calculate $F_c$ using Eq.(1)
one gets values shown by full line in Fig 9 which are found from Eq.(25) with 
$\chi^{-1}$=9.4 and $f_c$=4.1 pN. 

In conclusion, we developed a method to calculate the distribution of shear
force along the length of a dsDNA molecule at a finite temeprature.
The value of shear force $F_c$ which is required to separate 
the two strands of a molecule has been calculated
in both the constant force and the constant extension ensembles. The values
of $F_c$ found by these methods differ for molecules of shorter length 
but approach to each other as length increases. Both methods
gave the saturation value equal to 61.2 pN which is in very good agreement 
with the experimental value 61.4pN. The value of $F_c$ is found to increase
linearly with length up to 21 base pairs in agreement with experimental
results. The plots of $<u_n>$ and $<v_n>$ given in Fig.5 show that as long as 
the applied force pulls the entire strand in its direction of application,
$F_c$ depends linearly on length of molecules. The departure from
linear behaviour takes place when part of a strand moves in opposite
direction under the influence of  force pulling the other strand. The
saturation value is achieved when half of a strand moves in the
direction of force applied on it and the other half in the opposite
direction. It is shown that the value of $F_c$ quickly attains its 
saturation value on increasing length of DNA molecule; for a molecule 
of 32 base pairs $F_c$= 57.0pN as compared to saturation value 61.2pN.

The agreement between theorectical and experimental values of
$F_c$ shown in Fig.6 indicates that the model proposed in this paper
is capable of describing the responses of dsDNA molecules to 
shear stress. The model, however, neglects the effect of helicity
of DNA and has assumed the molecule to be homogeneous. As far as the effect 
of helicity is concerned, it has already been shown to be negligible 
by Hatch et al. \cite{hatch} and by Lavery and Lebrun \cite{lavery}. 
The effect of heterogenity arising due to random distribution 
of A-T and C-G base pairs in a sequence is also expected to be small
as long as the shear unzipping from the two ends is symmetric. This is  corroborated
by the fact that although the molecules investigated by
 Hatch et al \cite{hatch} are heterogeneous with half 
G-C and half A-T base pairs, yet their response to shear stress is described
very well by a model which assumes  molecules to be homogeneous.
The values of  parameters $\epsilon$ and $\kappa$ will, however,
depend on the percentage of A-T and C-G base pairs in a given sequence
and on the temperature.
But, in the absence of symmetry in the unzipping from the two ends a 
qualitative new features, as argued in ref \cite{chak}, may arise.
  
\subsection*{ACKNOWLEDGEMENTS }
We are grateful to Sanjay Kumar for drawing our attention to 
the paper of P. G. deGennes and to B. P. Mandal for useful discussions.
One of us (SP) acknowledges the financial support provided by the 
University Grants Commission, India.

\newpage
\begin{center}
\bf \large {Appendix}
\end{center}
In this appendix we derive an expression for $y_n$ from Eq.(9). Writing 
$v(y_n)$ as (see Eqs.(22) and (23))

\ba
v(y_n)=-\epsilon+\frac{1}{2}Ry_n^2   \nonumber \hspace{6cm} (A1)
\ea

where $R$=$\frac{12\epsilon}{\sigma^2}$ and differentiating with $y_n$ for 
$-\frac{N}{2}<n<\frac{N}{2}$ we get the following equilibrium conditions 
\cite{gennes} from Eq.(9)

\ba
-\frac{\partial H}{\partial y_n}=\kappa(y_{n+1}-2y_n+y_{n-1})-Ry_n=0 \nonumber \hspace{6cm} (A2)
\ea
For $n=N/2$, one gets
\ba
\frac{\partial H}{\partial y_{N/2}}=\kappa(y_{N/2}-y_{N/2-1})+Ry_{N/2}=\frac{F}{\sqrt2}
\nonumber \hspace{6cm} (A3)
\ea
From these equations we find 
\ba
y_n=\frac{1}{2}y_0({\it e}^{\chi n}+{\it e}^{-\chi n})
\nonumber \hspace{6cm} (A4)
\ea
where $\chi$=$\sqrt\frac{R}{\kappa}$,
and

\ba
y_0=\frac{F}{\sqrt2}\frac{1}{R{\it cosh}(\frac{1}{2}\chi N)[\chi^{-1}{\it tanh}(\frac{1}{2}\chi N)+1]} 
\nonumber \hspace{3cm} (A5)
\ea
Let the force on the last hydrogen bonds ($n$=$N/2$) is $R y_{N/2}$ when we reach 
the threshold $f_c$. ( $f_c$ being the rupture force for a base pair). Thus
\ba
f_c=Ry_0{\it cosh}(\frac{1}{2}\chi N) \nonumber \hspace{6cm} (A6)
\ea
From Eq.(A3) we get
\ba
F_c=\sqrt2\kappa\chi y_{N/2} {\it tanh}(\frac{1}{2}\chi N)+\sqrt2 f_c
\nonumber \hspace{6cm} (A7)
\ea
\ba
=\sqrt 2f_c[\chi^{-1}{\it tanh}(\frac{1}{2}\chi N)+1] \nonumber \hspace{6cm} (A8)
\ea
This equation differs from Eq.(1) in the definition of $\chi$ and the mutiplying 
factor which is now $\sqrt 2$ instead of 2.

\begin{figure}[ht]
\label{fig1}
\vspace{3.5cm}
\includegraphics[scale=0.5]{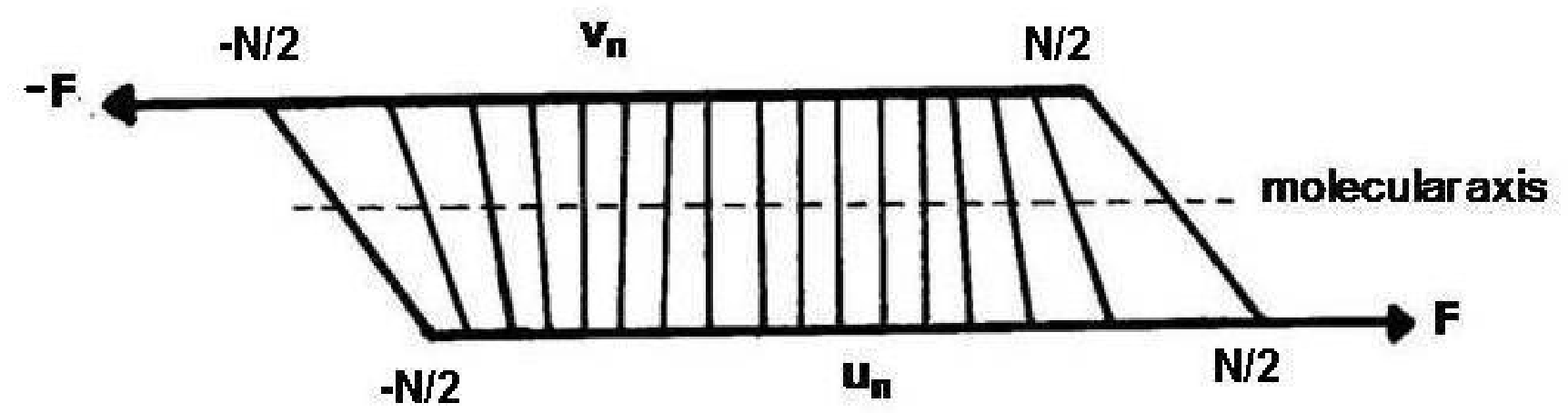}
\caption{\small Schematic of a dsDNA molecule of N+1 base pairs
 under a shear stress. $\bf u_n$ and $\bf v_n$ are displacements of
n$^{\it th}$ nucleotide in lower and upper strands, respectively.}
\vspace{0.1cm}
\end{figure}

\begin{figure}[hb]
\label{fig2}
\vspace{0.1cm}
\includegraphics[scale=0.8]{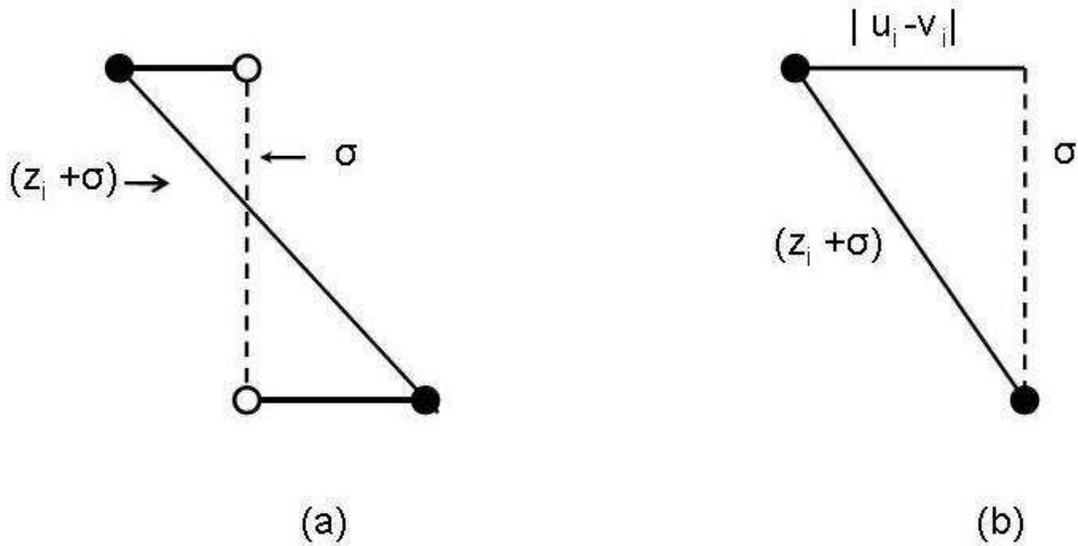}
\caption{\small Stretching of a base pair under the influence of shear
stress. The open circles in (a) indicate the equlibrium position
and filled circles stretched positions.}
\vspace{0.1cm}
\end{figure}

\begin{figure}[ht]
\label{fig3}
\vspace{1cm}
\includegraphics[scale=0.5]{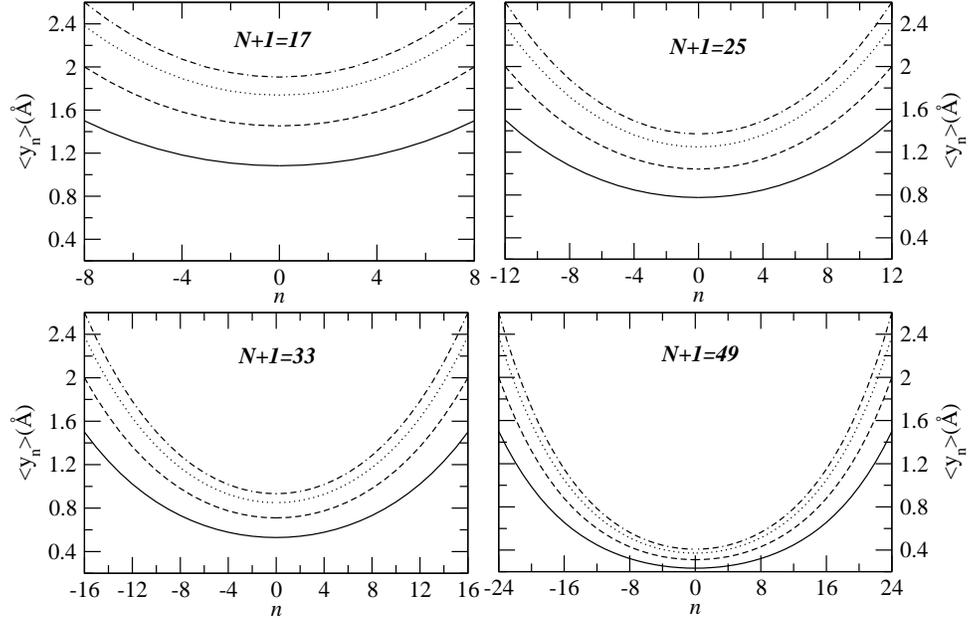}
\caption{\small Displacement $<y_n>$ as function of $n$ for four 
molecules of length 17, 25, 33 and 49 base pairs and for $y_{-N/2}=y_{N/2}$
=1.5$\rm \AA$ (full line), 2.0$\rm \AA$ (dashed line),2.38$\rm\AA$ (dotted line)
and 2.6$\rm\AA$ (dot-dashed line).}
\vspace{0.5cm}
\end{figure}

\begin{figure}[hb]
\label{fig4}
\vspace{1cm}
\includegraphics[scale=0.5]{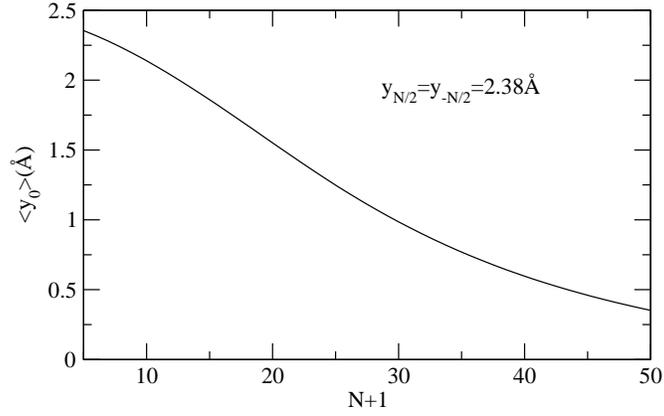}
\caption{\small Displacement $<y_0>$ of the central base pair as a function
of number of base pairs of dsDNA molecules.}
\vspace{0.5cm}
\end{figure}

\begin{figure}[ht]
\label{fig5}
\vspace{1cm}
\includegraphics[scale=0.5]{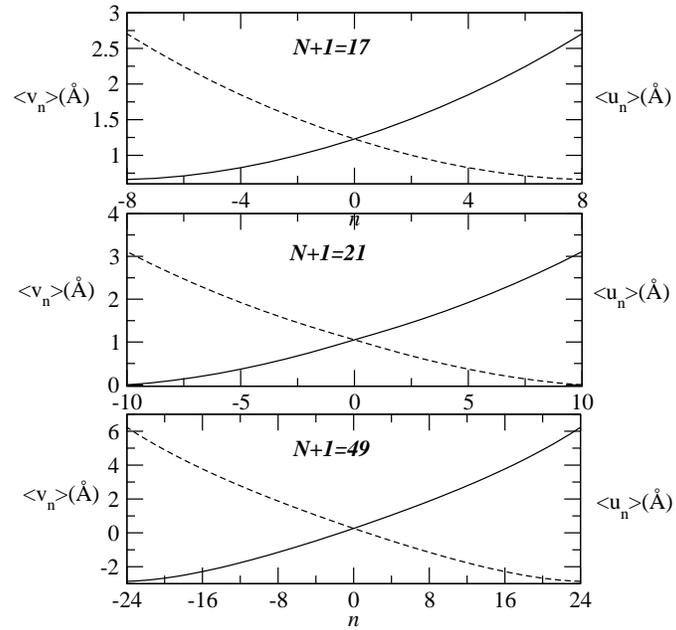}
\caption{\small The displacements $<u_n>$ and $<v_n>$ as a function
 of n of molecules of 17, 21  and 49 base pairs for the shear force $F_c$= 40.9, 47.4 
and 60.6 pN, respectively.}
\vspace{0.5cm}
\end{figure}

\begin{figure}[hb]
\label{fig6}
\vspace{1cm}
\includegraphics[scale=0.5]{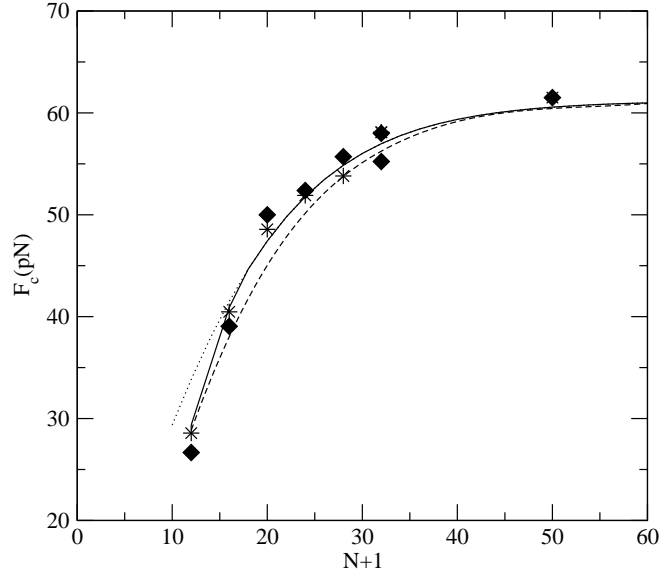}
\caption{\small The value of critical force $F_c$ in pN as a function
of number of base pairs of a dsDNA molecule. The curve shown by dotted line
corresponds to constant force ensemble and dashed line to constant
extension ensemble. The curve shown by full line corresponds to 
values found by taking the effective length 10 and 15 base pairs
for molecules of length 12 and 16 base pairs respectively.
The experimental values of Hatch et al \cite{hatch} are shown by
diamond when 5'-ends are pulled and by star when 3'-ends are pulled
by shear force.}
\vspace{0.5cm}
\end{figure}

\begin{figure}[ht]
\label{fig7}
\vspace{1cm}
\includegraphics[scale=0.5]{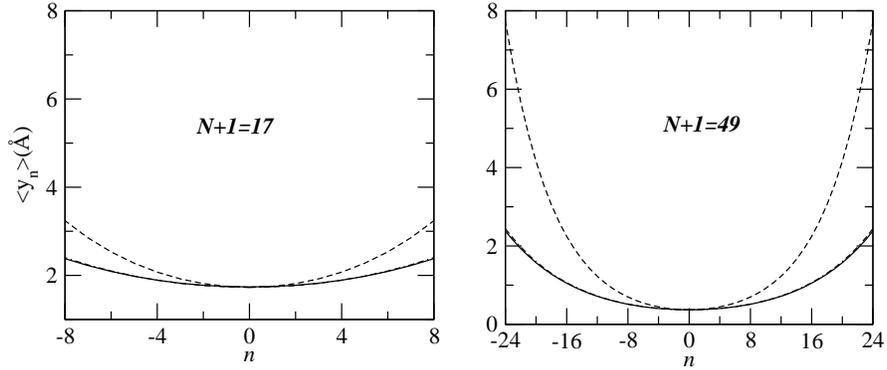}
\caption{\small Comparison of the values of $<y_n>$ found from
Eq.(12) (full line) and the deGennes relation Eq.(24) (dashed line)
when $\chi^{-1}$=6.44. For $\chi^{-1}$=9.34, the values found from Eq.(24)
overlap with the curve drawn by full line.  }
\vspace{0.5cm}
\end{figure}

\begin{figure}[hb]
\label{fig8}
\vspace{1cm}
\includegraphics[scale=0.5]{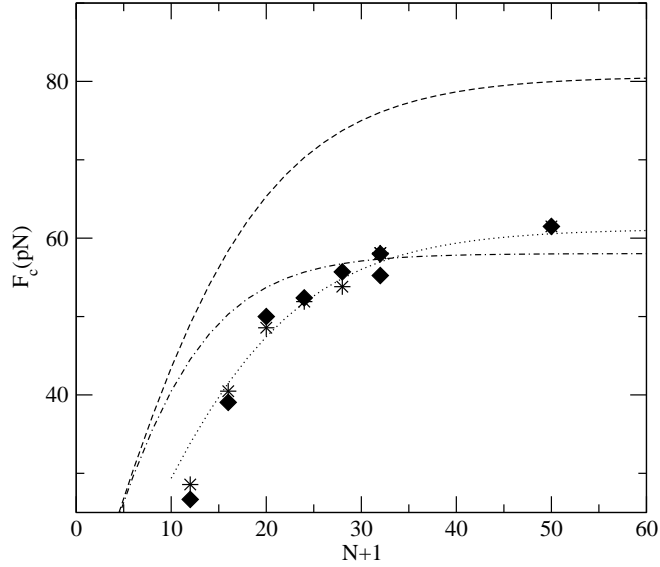}
\caption{\small Comparison of the values of shear force $F_c$ as a function
of the number of base pairs in a molecule. The curves shown by dashed and 
dot-dashed lines represent
the values found from  Eq.(25) when $\chi^{-1}$=9.34 and 6.44, 
respectively. The dotted line  represents corresponding curve of Fig.6.}
\vspace{0.5cm}
\end{figure}

\begin{figure}[htp]
\label{fig9}
\vspace{1cm}
\includegraphics[scale=0.5]{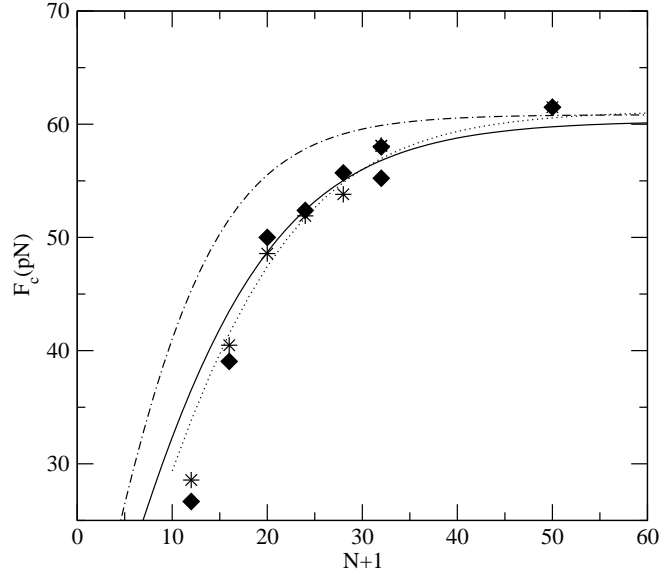}
\caption{\small Comparison of the value of shear force $F_c$ as a function of
the number of base pairs in a molecule. The curve shown by dot-dashed line is 
found from Eq.(1) with $\chi^{-1}$=6.8 and $f_c$=3.9pN \cite{hatch}, the 
curve shown by full line is found from Eq.(25) with $\chi^{-1}$=9.4 and
$f_c$=4.1pN and the dotted line represents corresponding curve of Fig.6.}
\vspace{0.5cm}
\end{figure}
\end{document}